\documentclass[aps,prl,twocolumn,showpacs,groupedaddress]{revtex4}
\usepackage{amsmath}
\usepackage{amssymb}
\usepackage{graphicx}
\newcommand{\um}{\hbox{\,$\mu$m}}
\newcommand{\mys}{\hbox{\,s}}
\begin{document}

\title{Nonlinear self-adapting wave patterns}
\author{David A. Kessler}
\affiliation{Department of Physics, \\Bar-Ilan University, Ramat Gan, IL52900 Israel}
\author{Herbert Levine}
\affiliation{Center for Theoretical Biological Physics, \\Rice University, Houston TX, 77251-1892}
\date{\today}
\begin{abstract}
 We propose a new type of traveling wave pattern, one that can adapt to the size of physical system in which it is embedded. Such a system arises when the initial state has an instability that extends down to zero wavevector, connecting at that point to two symmetry modes of the underlying dynamical system. The Min system of proteins in \textit{E. coli} is such as system with the symmetry emerging from the global conservation of two proteins, MinD and MinE. For this and related systems, traveling waves can adiabatically deform as the system is increased in size without the increase in node number that would be expected for an oscillatory version of a Turing instability containing an allowed wavenumber band with a finite minimum. 
\end{abstract}


\maketitle

One of the ways in which a non-equilibrium system can lead to pattern formation is via a traveling wave bifurcation`\cite{cross}. In such a system, the uniform state becomes unstable to modes at finite wave vector $k$ and finite frequency $\omega$ leading to a variety of phenomena involving nonlinear traveling wave states.  This scenario has proven relevant for processes ranging from binary fluid convection~\cite{binary1,binary2} to electro-hydrodynamics in liquid crystals~\cite{ehd} to the sloshing of Min proteins in bacteria~\cite{min1,meinhardt,wingreen,frey1,Kruse,kerr}

To date, this instability has been viewed as an oscillatory analog of the familiar Turing instability; this means that the instability occurs at fixed $q \neq 0$,  and above threshold the unstable band stretches from $0 < q_{min} < q < q_{max}$. Here we show that there exists a new possibility, namely that for some systems with the correct symmetry, $q_{min}$ equals zero.  This dramatically changes the nature of the nonlinear patterns that form, as there is no predetermined length scale for the emergent structure; instead, the waves are able to self-adapt to the size of the physical system. As we will discuss, this is the traveling wave analog of what happens in viscous fingering~\cite{kkl,viscous} where the {\em static} bifurcation extends to $q_{min}=0$. Models for the aforementioned Min dynamics offer a specific realization of this new paradigm. Moreover, the self-adaptation provides a mechanism whereby the dynamical pattern can maintain a one node form as  the cell expands during growth.

We start with the model of Ref. \cite{Kruse,kruse2} for the Min system. There are two proteins, MinD and MinE, each of which can be on the membrane ($m$) or in the cytosol ($c$).  The two proteins can reversibly desorb and adsorb, and adsorption of MinE involves it directly binding to an already membrane resident MinD. Additional nonlinearities emerge from the assumed cooperativity in the desorption rates. Adding in diffusion in the  compartments leads in one spatial dimensions to the 4 coupled pde's
\begin{align}
\frac{\partial c_D}{\partial t} & = R_{de\to D+E} - R_{D\to d}  +D_D \frac{\partial ^2 c_D}{\partial x^2} \nonumber\\
\frac{\partial c_E}{\partial t} & =  R_{de\to D+E} - R_{d+E\to de} +D_E \frac{\partial ^2 c_E}{\partial x^2} \nonumber\\
\frac{\partial c_d}{\partial t} & =  R_{D\to d} - R_{d+E\to de}  +D_d \frac{\partial ^2 c_d}{\partial x^2} \nonumber\\
\frac{\partial c_{de}}{\partial t} & =  R_{d+E\to de} - R_{de\to D+E} +
D_{de} \frac{\partial ^2 c_{de}}{\partial x^2}
\end{align}
where $c_D$ and $c_E$ are the cytosol concentrations of MinD and MinE, $c_d$ is the concentration of MinD on the membrane and $c_{de}$ is the concentration of the MinD/MinE complex on the membrane and the rates are
\begin{align}
R_{D\to d} &= (\lambda_D + \lambda_{dD}c_d) c_D \nonumber\\
R_{de\to D+E} &= \lambda_{de}c_{de} \nonumber\\
R_{d+E\to de} &= (\lambda_E + \lambda_{eE}c_{de}^2)c_dc_E \nonumber\\
\end{align}
 \begin{figure}[t]
 \includegraphics[width=0.9\linewidth,clip=]{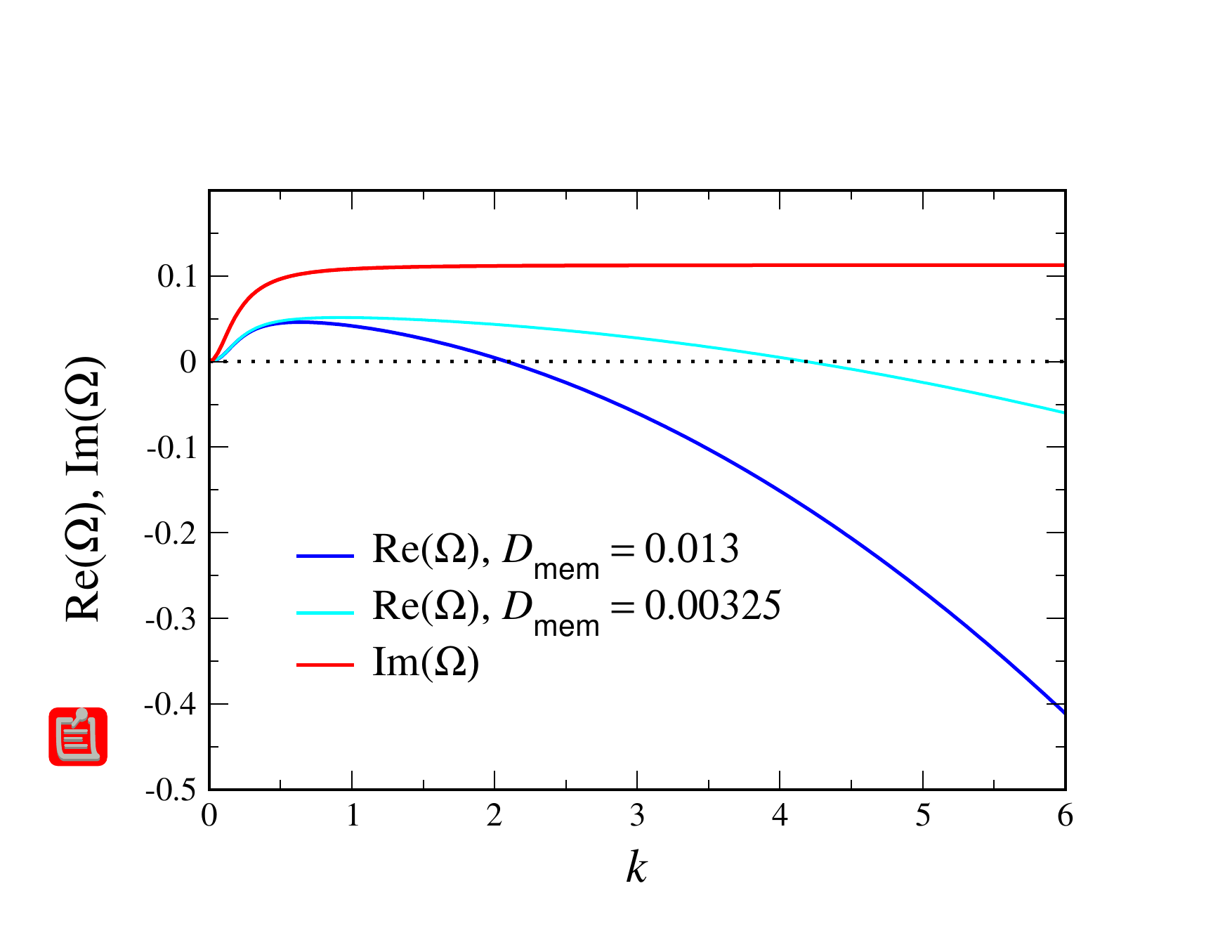}
 \caption{ The growth rate $\textrm{Re}(\Omega)$ and frequency $\textrm{Im}(\Omega)$ of the most unstable mode of the linear stability operator constructed about the uniform steady-state solution, for two different values of the diffusion rate, $D_m$, of the two membrane-bound species, $D_m=D_d = D_{de}$. The frequencies for the two cases are indistinguishable at the resolution of the graph. The other parameters, as taken from Ref. (\cite{Kruse}), are: $D_D=D_E=12.5\um^{2}\mys^{-1}$, $\lambda_D=0.0013\mys^{-1} $, $\lambda_{de}=0.125\mys^{-1}$, $\lambda_{dD}=9.3\cdot 10^{-4}\um\mys^{-1}$, $\lambda_E=3.8 \cdot 10^{-5}\um\mys^{-1}$, $\lambda_{eE}=8 \cdot 10^{-9}\um^3\mys^{-1}$, $D_T=1000\um^{-1}$, $E_T=400\um^{-1}$. In the following, all lengths with be in $\!\um$ and times in $\!\mys$ and explicit units will be dropped.}
 \end{figure}
Of critical importance to the physics of this system are two features. First, the diffusion constants of the membrane-bound species are orders of magnitude smaller than their values for the same protein in the cytosol. We will see that this property will guarantee the existence of an instability which drives the pattern formation and also will enable a simplification of the dynamics, at least for small system size. Second, the model contains two global conservation laws; the total number of both MinD and MinE proteins in the system are unchanged by each of the reactions, so that they are determined solely by the initial conditions. To see what this implies, we imagine we have found a uniform steady-state solution of the model and calculate the fate of  a small perturbation around this solution, so that $$c_D = c_D ^{(0)} +\delta_D e^{-ikx +\Omega t}$$ with analogous expressions for the other concentrations. This leads in a standard manner to a $4\times 4$ homogeneous linear system which determines the allowed eigenvalues. 
The results of such a calculation for the most unstable mode are presented in Fig. 1.  We see that for $k<k_{*}$, there is a complex unstable mode, whose (complex) growth rate goes to 0 in the limit $k\to 0$. We shall now show that this is a direct result of the global conservation laws.  At $k=0$, due to these   laws, the sums of equations \{1,3,4\} and equations \{2,4\} are identically zero, as  shifts in the overall levels of either the conserved MinD or MinE proteins due to the perturbation are left unchanged by the dynamics. Hence $\Omega=0$ is a  doubly degenerate eigenvalue for $k=0$, with left eigenvectors $\hat{\phi} ^{(1)} =(1,0,1,1)$ and $\hat{\phi} ^{(2)} =(0,1,0,1)$ . We can then evaluate the effect of non-zero $k$ to leading order by simply computing the projection of the diffusion terms along the diagonal, namely ${\mathcal D} = diag( -D_D, -D_E,-D_d,-D_{de}) k^2 $ onto the basis of the degenerate $2x2$  subspace to obtain
$$\mathcal{S}_{ij} = {\hat \phi ^{(i)}} {\mathcal D} \phi^{(j)} $$
where the $\phi^{(j)}$ are the corresponding right eigenvectors which satisfy the orthonomality condition $\hat{\phi} ^{(i)} \phi ^{(j)} = \delta_{ij}$. For the parameter set used in \cite{Kruse}, the steady-state is $c_D^0=  71.77$, $c_E^0=76.39$, $c_d^0=604.62$, $c_{de}^0=  323.61$    and this gives rise to 
\[ 
\phi^{(1)} = \left( \begin{array}{c}
-0.0767 \\ -0.1494 \\ 0.9273 \\ 0.1494 \end{array} \right) \qquad \phi^{(2)} = \left( \begin{array}{c}
0.4022 \\ -0.0422 \\ -1.4444 \\ 1.0422 \end{array} \right) 
\]
\[
   \mathcal{-S}/k^2 = 
 D_c \left[ {\begin{array}{cc}
   -0.077 & 0.402 \\      -0.149 & -0.042 \      \end{array} } \right]   +  D_m \left[ {\begin{array}{cc}
   1.077 & -0.402 \\       0.149 & 1.042 \      \end{array} } \right] 
\]
for pair-wise equal cytosol ($D_c$) and membrane ($D_m$) diffusivities. This matrix always has a pair of complex eigenvalues, as
\begin{equation}
\textrm{tr}^2(S)-4\textrm{det}(S) =  -0.239 k^4 (D_c-D_m)^2 < 0
\end{equation}
This complex pair is unstable as long as 
\begin{equation}
\textrm{tr}(S) = k^2 \left( -0.119 D_c  + 2.119 D_m \right) < 0 \Rightarrow \frac{D_c}{D_m} > 17.8
\end{equation}
which is easily satisfied by the biophysical parameters, as we saw in Fig. 1. In general, the initial rise of $Re \Omega$ with $k$ depends on the relatively large cytosol diffusion constants whereas the  value of $k$ at which the system restabilizes depends on the small membranal ones.  In the limit of very large diffusion constant ratio between the cytosol and membranal fields and for equal  membranal diffusivities, $D_m$, the spectrum approaches (for $k$ strictly non-zero) the simple form $\Omega _0 - D_m k^2$ for complex $\Omega_0$ with a positive real part.

This stability structure presents a new twist on what happens in pattern forming systems such as viscous fingering and dendritic crystal growth~\cite{kkl,viscous}. There, translation invariance of the base system guarantees a single zero $k=0$ eigenvalue which gives rise to a real-mode instability for $0<k<k^*$. A related idea has arisen in the context of cellular processes that have one chemical component being exchanged between different compartments but is globally conserved \cite{wave-pinning}. In our system the existence of two zero modes and of course the non-symmetric nature of the stability matrix allows for a pair of complex conjugate modes to have a positive growth rate. The study of those interfacial systems has revealed characteristic differences between the nonlinear states that emerge as compared to those in related systems such as directional solidification~\cite{langer} which have a regular Turing-like mode spectrum.  Here the basic pattern is the traveling wave, which due to the instability extending down to $k=0$, should exist at very large wavelengths.  We now turn to a study of this pattern.
\begin{figure}
\includegraphics[width=0.45\textwidth]{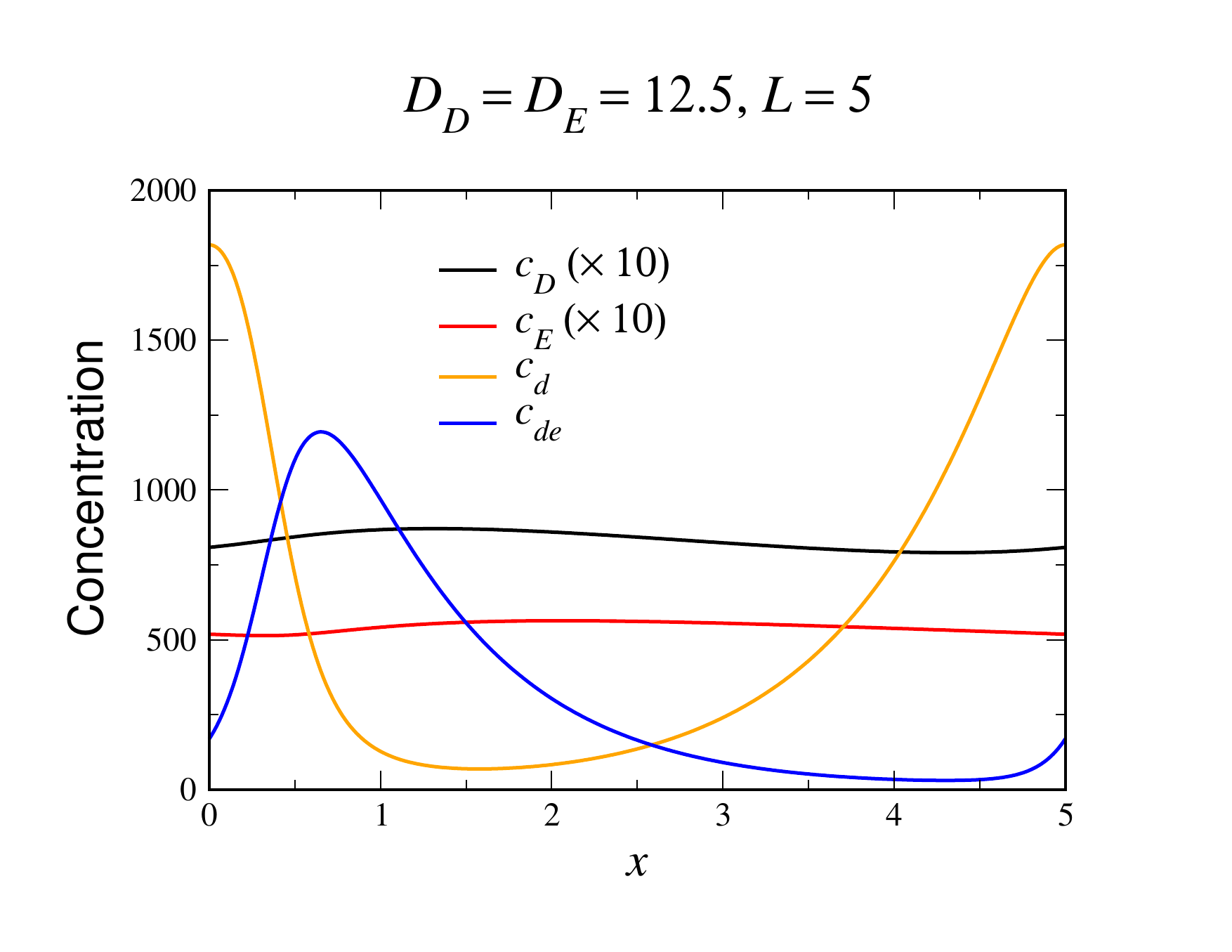}\\\includegraphics[width=0.45\textwidth]{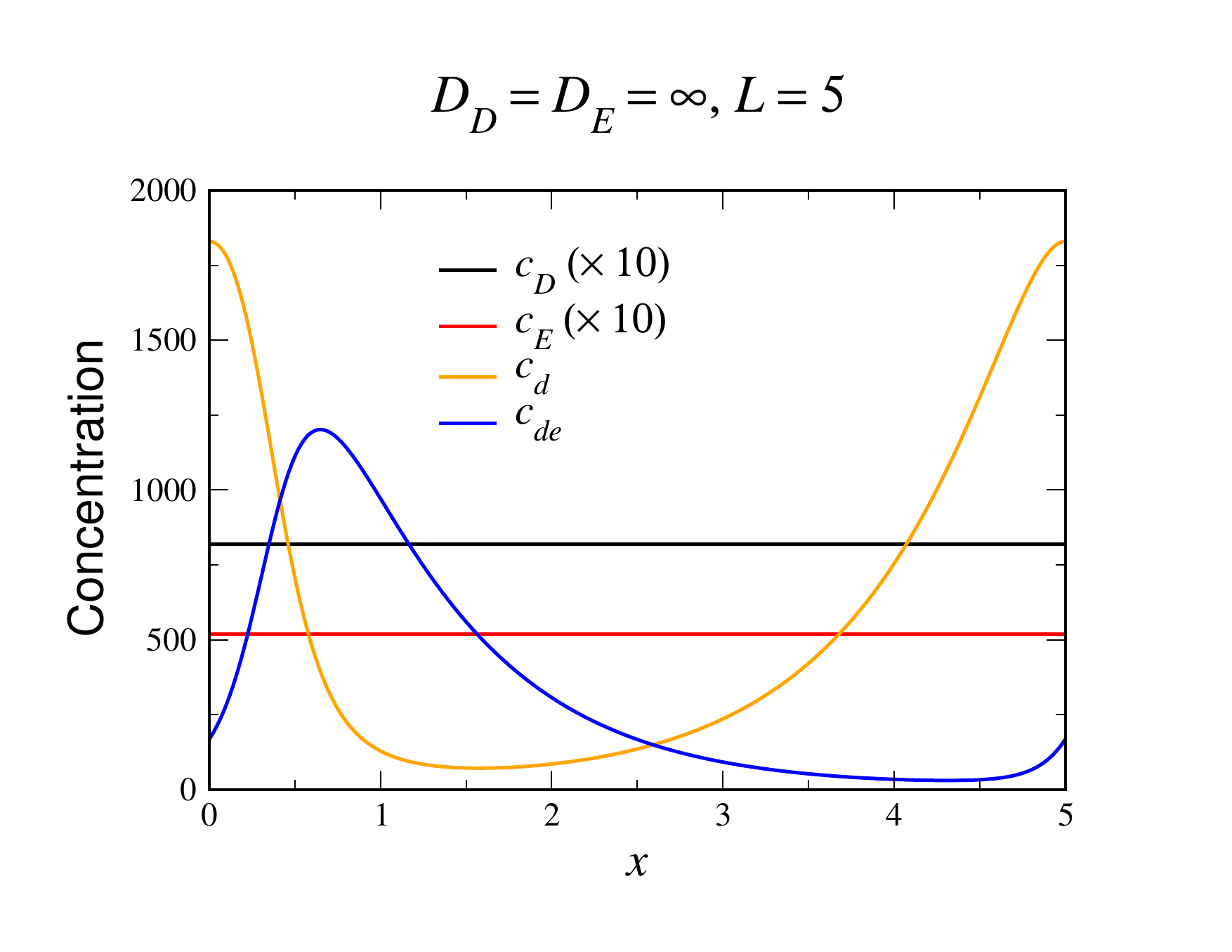}\\\includegraphics[width=0.45\textwidth]{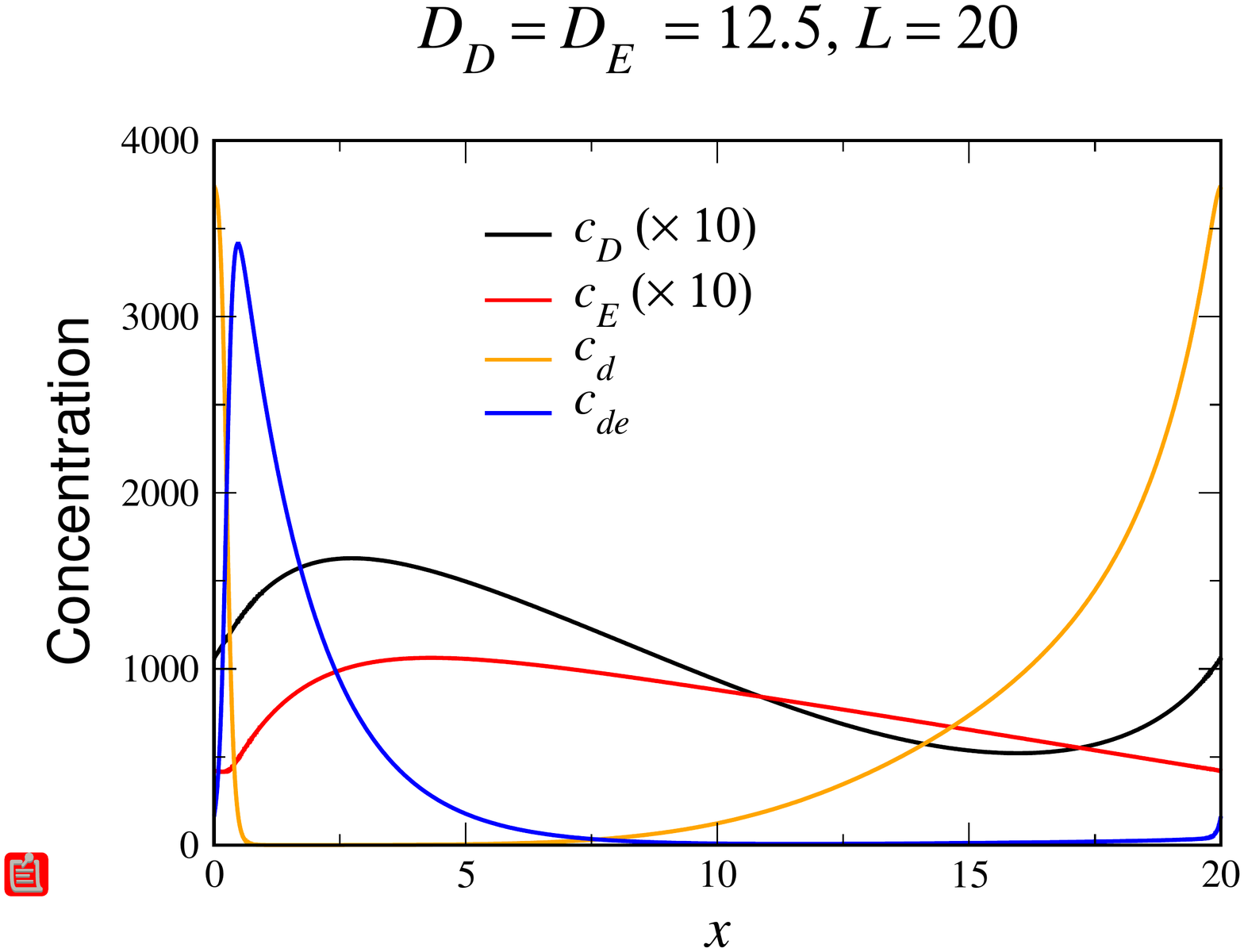}
\caption{The uniformly propagating wave solution. Top: The solution for the parameters of Ref. \cite{Kruse}, for which the stability analysis is shown in Fig. 1 (blue curve).  The length of the periodic system is $L=5$.  Middle:  The solution for the limit $D_c = \infty$, with all other parameters as above.  The membranal field profiles are basically unchanged from the above graph.  Bottom: The solution for the same parameters as in the top panel, for the larger system $L=20$.  The peak in the membranal fields is roughly the same width as in the top panel, so that  $L$-scaled coordinates, it appears much sharper.  Away from the peak the solutal fields appear similar to the top panel, indicating that these features scale linearly with $L$. The variation of the solutal fields is much increased over that of the top panel.}
\label{fig2}
\end{figure}

In the top panel of Fig 2, we show an example of a traveling wave solution, corresponding to the parameter set already used above, for a periodic system of size $L=5 \mu$. The second  takes the limit of infinite diffusivity for the cytosolic species $c_D$ and $c_E$; for the latter case, the model is globally coupled with the value of these fields determined at all times by the integral constraints
\begin{eqnarray*}
L c_D & = & D_T - \int  dx \left( c_{d} +c_{de} \right) \\
L c_E & = & E_T - \int  dx \ c_{de} 
\end{eqnarray*}
where $L$ is the size of the periodic domain. For this size system, which is not much larger than the minimal size for the instability, $L_{min}=2\pi/k^*\approx 3.0$, this traveling wave pattern appears to be the unique attractor of the system, arising from generic initial conditions. The solution can thus be generated by running a simulation and waiting for the system to settle into this uniformly propagating state, which perforce must be linearly stable. Alternatively, we can directly solve the steady-state equation in the moving frame of reference by an iterative scheme acting upon the field values at collocation points.  Since there are 4 second-order equations, we have to impose eight conditions.  Six of these are the continuity
of the four fields and two of the derivative fields across $x=L$.  Two are the global constraints on the $D_T$ and $E_T$. Because of translation invariance we can arbitrarily choose one of the fields to have a known value at say $x-vt =0$ and reduce the number of unknowns by one. Then the number of equations to be solved in one greater than the field unknowns, necessitating the use of the velocity as the final unknown. One can check that we get the same results from both of these methods. The second approach is specifically convenient if one has a solution for some parameter set and wishes to find a solution at a nearby one, as in that case there is a very good initial guess with which to start the iteration.   We can see from these graphs that there is nothing singular about the infinite-$D_c$ global limit at least as far as this type of solution is concerned. 

We see that while the cytosol concentrations are relatively featureless (exactly so in the $D_c \to \infty$ limit), the membranal fields each have a single peak, located close to each other.  As we take $L$ larger, as in the bottom plot of Fig. 2, this structure is maintained, with the peaks having roughly the same width, and so occupying a smaller fraction of the system. The system settles into a scaling form of the solution in which the pattern consists of two parts. There is an inner region, which for Fig 2c lies at around $x/L =$ which gets thinner (in rescaled coordinates) as $L$ increases. The rest of the box has an ``outer" solution which scales linearly with $L$. The velocity of the solution scales as $L$ once the system is in the scaling regime (data not shown), which for our parameters occurs for $L \gtrsim 15 $.  It is interesting to note that the large cytosol diffusion is much less successful in eliminating the variation in the cytosol fields in the larger $L$ system.

We can understand this solution by looking separately at the two aforementioned regions, in the globally coupled limit, keeping $D$ and $E$ fixed as we increase $L$. We assume a dependence only on $z=x-vt$ and hence the time derivatives become $-v \frac{d}{dz}$. In the outer region, the slow diffusion is irrelevant and the only spatial derivative is the velocity term. Thus, having $v \sim L$ immediately allows the outer solution to have spatial decays away from the peak which become $L$ independent in the rescaled coordinate. In the inner region, we have in general three terms that are important; the velocity term, the diffusion term and the cubic term that occurs (with opposite sign) in both the $c_d$ and $c_{de}$ equations. In fact, if we add the two equations and assume equal diffusivities, we get that $c_d +c_{de} $ does not  have any driving term and one can easily check from the numerical solution that it is a smooth function on the inner scale. Let us denote by $A$ the constant value of this sum at the location of the inner zone. If we rescale lengths by $z=\tilde{z} /L$, velocity by $v =\tilde{v} L$, we get the equations
\begin{eqnarray*}
0 & = & \tilde{v} \frac{dc_d}{dz} + D_m \frac{d ^2 \tilde{c}_d}{d z^2}   - \frac{\lambda_{eE}}{L^2} c_E  \tilde{c}_d  (A- \tilde{c}_{d})^2 \\
0 & = & \tilde{v} \frac{dc_{de}}{dz} + D_m  \frac{d ^2 \tilde{c}_d}{d z^2}   + \frac{\lambda_{eE}}{L^2} c_E  \tilde{c}^2_{de}  (A- \tilde{c}_{de}) 
\end{eqnarray*}
This pair of conjugate equations are familiar from the literature on pattern formation. We can define a ``potential" function $U(c_d)$ to recast the equation for $c_d$, e.g. as that of a "particle" moving a well
$$ -\frac{dU}{dz}  =  \tilde{v} \frac{dc_d}{dz} + D_m \frac{d ^2 \tilde{c}_d}{d z^2}  $$ where the potential is obviously 
$$U(c_d )  = -\frac{\lambda _{eE}}{L^2} \left( A^2 c_d^2/2 -2A c_d^3/3  +c_d^4/4 \right)$$. 
The inner solution is then a particle which rolls as $z$ goes from $-\infty$ to $+\infty$ from the $U$ maximum at $c_d =0$ to the $U$ point of inflection at $c_d=A$. From this analogy, it is obvious that solutions of this form exist for all velocities above a critical velocity which can be found to equal $\frac{A \ \sqrt{2 \lambda_{eE} D_m} }{L}$ by directly  substituting in the ansatz $c_d = \frac{A}{2} (1+\tanh (z/w))$. Note that for velocities above critical, there is actually a power-law decay to the fixed point at $A$ rather than the exponential decay obtained for the minimal velocity.   If we call this solution $\psi (z;v,A)$, we have the final forms
$c_d = \psi(z-z_{inner}; \tilde{v}, A)$ and $c_{de} = A-c_d$. There are two unknowns, namely $A$ and the scaled velocity $\tilde{v}$.

The construction is then completed by integrating the outer equations, where diffusion is ignored and lengths are now scaled as $ (z-z_{inner}) = \tilde{z} L$; this choice cancels the $L$ dependence in the velocity term and gives us an equation where all terms are ${\cal{O}}(1)$ and so $L$ has completely disappeared from the problem. One can then integrate the coupled first-order outer equations starting immediately past $z_{inner}$ with the initial conditions $c_d = A$, $c_{de} =0$ and demand that the solution at $\tilde{z} =1$ returns back to the inner solution left asymptote $c_d=0$, $c_{de}=A$. These conditions determine the two unknowns. The only possible difficulty is that the determined velocity $\tilde{v}$ could fall below the minimum velocity of the inner solution; since that latter scales as $1/L$ this will always occur as $L$ is decreased and in fact defines the lower $L$ limit for the existence of this self-adjusting wave. Waves do continue to exist below this $L$, but they have a more complex scaling indicated for example by the amplitude $A$ changing with $L$. The fact that the system supports a scale-invariant nonlinear traveling wave is, we believe, traceable to the nature of the original stability; the system can use its ability to self-amplify at any non-zero $q$ to form this solution.  

 \begin{figure}[t]
 \includegraphics[width=0.9\linewidth,clip=]{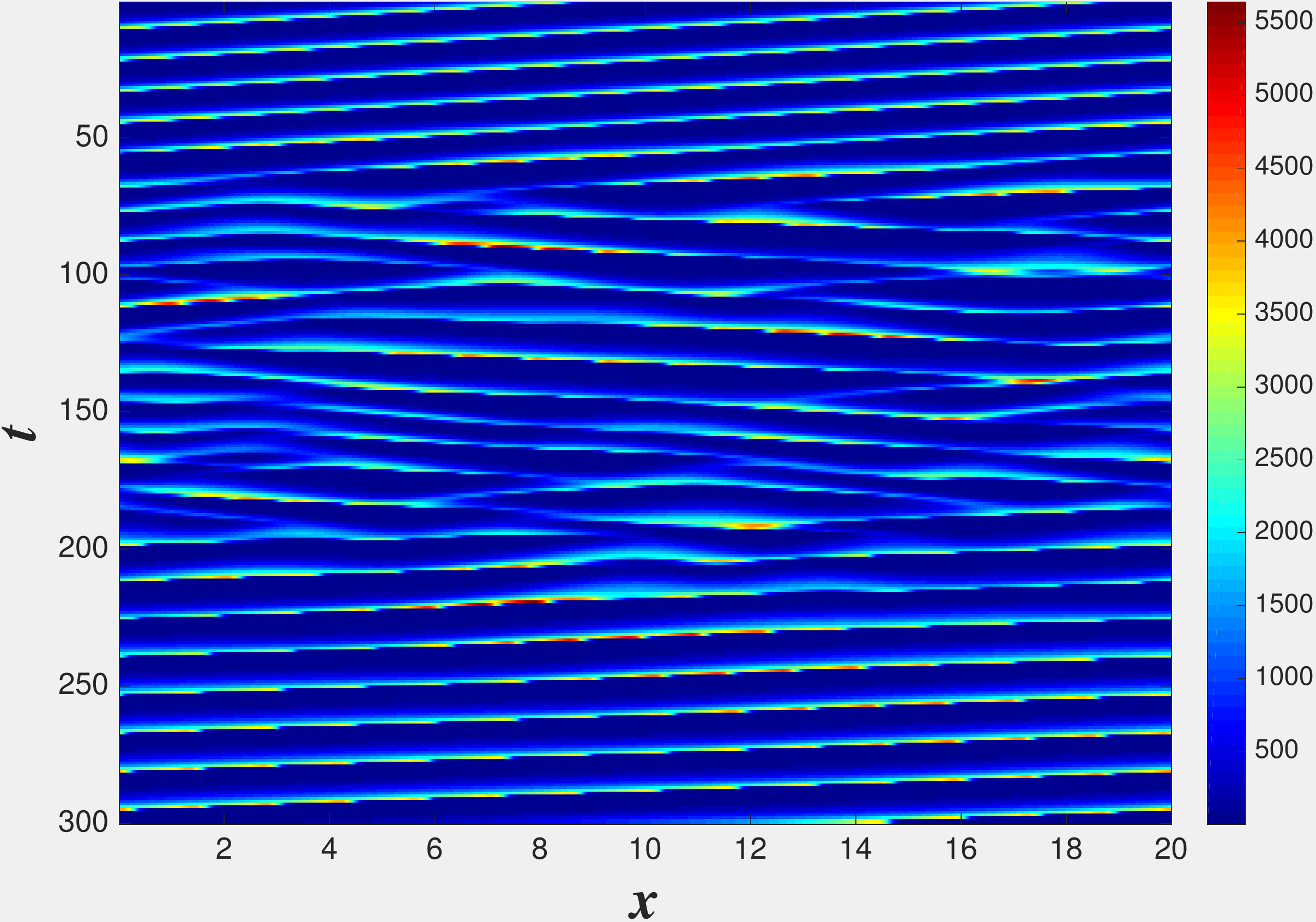}
 \caption{The merging of a two peak traveling wave composed of two $L=10$ solutions into a single $L=20$ steady traveling wave.
 The $c_d$ field is presented, the $c_{de}$ field would look very similar.  The parameters are the same as in the top panel of Fig. 2.}
 \end{figure}

While  this ``single pulse" wave appears to be the unique stable steady-state solution for relatively small $L$, as $L$ increases this ceases to be the case.
One way to see this is to start with an initial condition which is composed of two $L/2$ pulses.  For $L\lesssim 25$, for our ``standard" parameters,  the two peak solution develops an instability and eventually reaches the single peak solution appropriate to a system size of $L$. This process is demonstrated in Fig 3 for $L=20$.  This is analogous to what was established long ago for a periodic array of Saffman-Taylor fingers~\cite{coalescence}. But, here, the full story is complicated and in general we find three regions for the asymptotic state arising from these conditions. For $L <L_{c1} (D_m) $, we obtain full coarsening as above; at larger $L$ the two pulse pattern is stable, coexisting with the stable one pulse wave. At even larger $L$, there is a supercritical pitchfork bifurcation to a non-symmetric two pulse state. 
Analogously, one can start with a three pulse initial condition in a $3L$ box and find regions of stable non-symmetric solutions at large enough $L$. In addition, the larger range of unstable wavevectors at increasing $L$ appears to result in a shrinking of the basin of attraction of these steady-state solutions; this remains to be quantitatively analyzed. 

Because of the stability of the single pulse solution, the system will stay in this state as the box size is adiabatically increased. To see this, we assume that the system is regulated so as to maintain a fixed overall average concentration of MinD and MinE and insert more of these proteins uniformly into the cytoplasm as we expand the cell. Following Ref. \cite{expand}. the system equations then read
\begin{align}
\frac{\partial c_D}{\partial t} & =   \frac{\dot{L}(t)}{L(t)} (D_T - c_D) + R_{de\to D+E} - R_{D\to d} + \frac{D_D}{L^2(t)}  \frac{\partial ^2 c_D}{\partial y^2} \nonumber\\
\frac{\partial c_E}{\partial t} & =   \frac{\dot{L}(t)}{L(t)} (E_T - c_E) +  R_{de\to D+E} - R_{d+E\to de} +\frac{D_E}{L^2(t)} \frac{\partial ^2 c_E}{\partial y^2} \nonumber\\
\frac{\partial c_d}{\partial t} & =  - \frac{\dot{L}(t)}{L(t)} c_d +  R_{D\to d} - R_{d+E\to de}  +\frac{D_d}{L^2(t)} \frac{\partial ^2 c_d}{\partial y^2}\nonumber \\
\frac{\partial c_{de}}{\partial t} & =  - \frac{\dot{L}(t)}{L(t)} c_{de} +  R_{d+E\to de} - R_{de\to D+E} +
\frac{D_{de}}{L^2(t)} \frac{\partial ^2 c_{de}}{\partial y^2}
\end{align}
where $L(t)=L(0)e^{\gamma t}$ is the time-dependent length of the system and $y\equiv x/L(t)$ is the scaled spatial coordinate.
Fig. 4 shows clearly that the one pulse wave will maintain its global topology for a very large range of scales.  Of course, the actual Min system does not live in a periodic domain; even if one adopts the simplification of ignoring the actual compartment structure of the cell into membrane and cytosol in favor of a bi-continuous approach (as is done here), one should obviously use zero flux conditions at the cell edges. So, it is useful to ask about the from of the nonlinear traveling wave state to the dynamics in a fixed box. In the top panel of Fig. 5, we present snapshots of simulations for small cells, showing clearly a  ``sloshing'' wave  pattern with sharply decreased amplitude at the cell center. Most importantly this topology is not changed as the cell expands, even as the pattern becomes more like a traveling wave bouncing back and forth (see the bottom panel of Fig. 5). The time-average concentrations maintain a single node at the center even as the cell doubles; this is necessary for the functional role of the Min system in defining the precise midpoint of the cell~\cite{min2,kerr}. The self-adjustment property of the system allows this to take place without any fine-tuning of system parameters. 

 \begin{figure}[t]
 \includegraphics[width=0.9\linewidth]{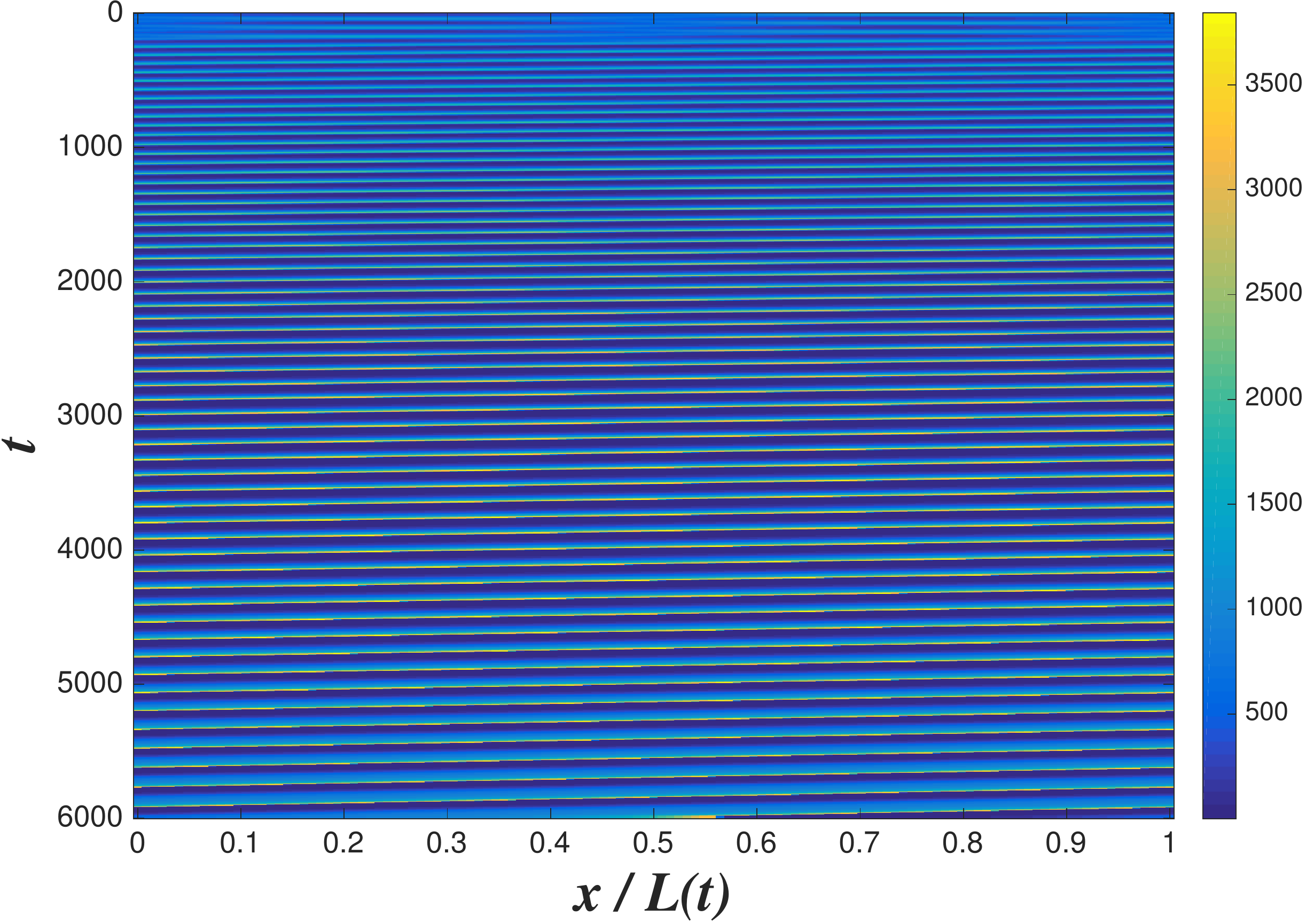}
 \caption{ The pattern produced by an exponentially growing system, with doubling time of 30 mins and initial size of $L=4$. Solutal MinD, MinE is added to keep the overall average MinD and MinE concentrations fixed. The membranal field $c_d$ is displayed, $c_{de}$ looks very similar. The system quickly settles into a single pulse traveling wave solution and preserves this form, despite the continually change in $L$. The increase in amplitude and the initial decrease in period are apparent. The other parameters are the same as in the top panel of Fig. 2.}
 \end{figure}

It is critical to realize that very few of our findings should have anything to do with the detailed assumptions of the model. For example, if one uses more recent and presumably more realistic models for Min dynamics proposed in refs. \cite{frey1,frey2}, the existence of two conservation laws will again guarantee that the wave instability will extend down to $q=0$ and therefore we can predict the existence of self-adjusting traveling wave states. This of course needs to be investigated in detail. A more uncertain situation holds for a recently studied case of a $q_{min}=0$ wave instability arising during the frictional sliding of one surface above a second~\cite{friction}. Here the fact that the base state with uniform sliding is explicitly not reflection symmetric and hence there need not be modes at both $+q$ and $-q$ at the same complex value of $\Omega$; in other words, there is a preferred direction of wave propagation and this one unstable wave can be connected to just one symmetry mode as $q \rightarrow 0$. The extent to which this difference matters for the non-linear state remains to be studied.

 \begin{figure}[t]
 \includegraphics[width=0.9\linewidth]{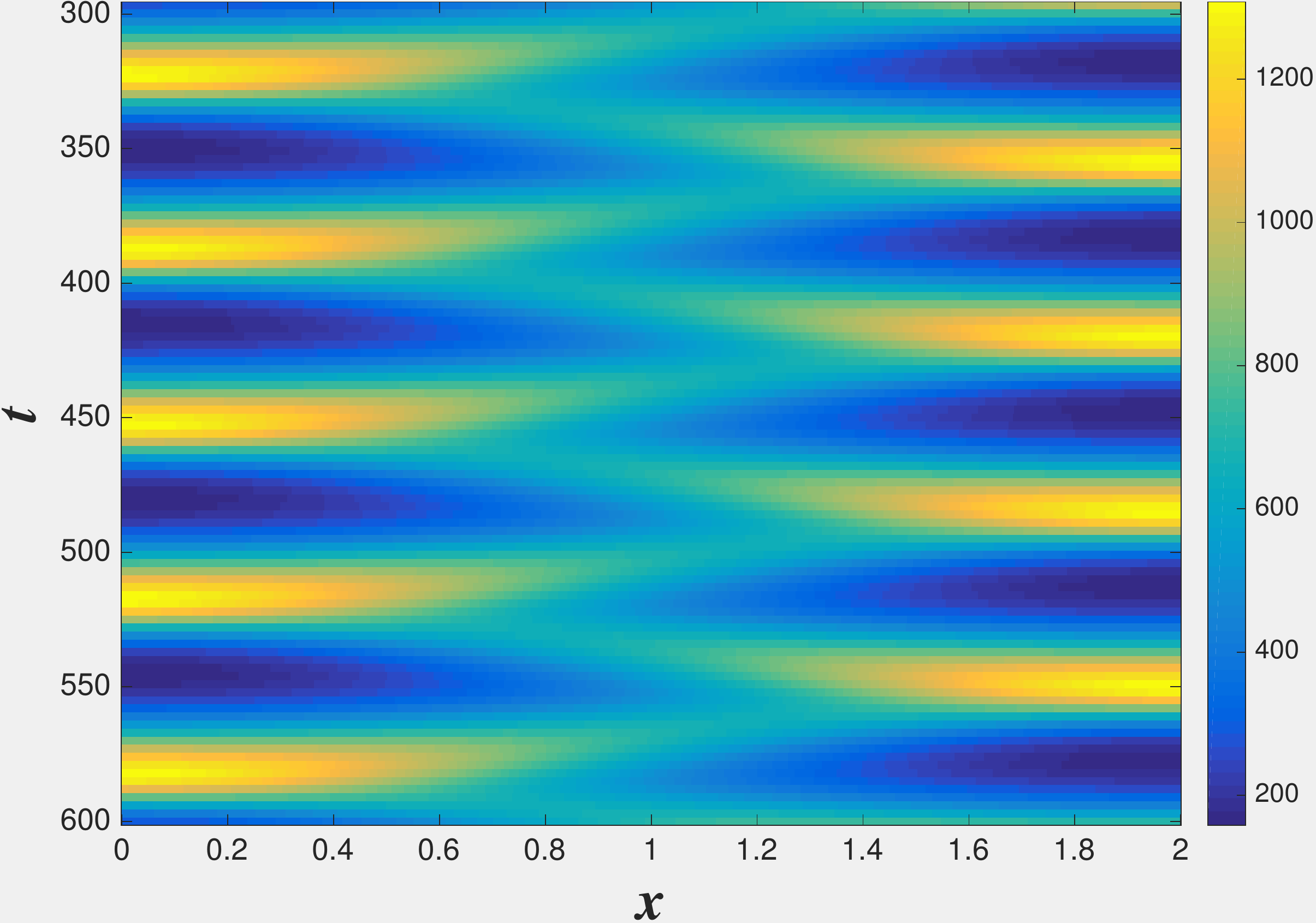}\\\includegraphics[width=0.9\linewidth]{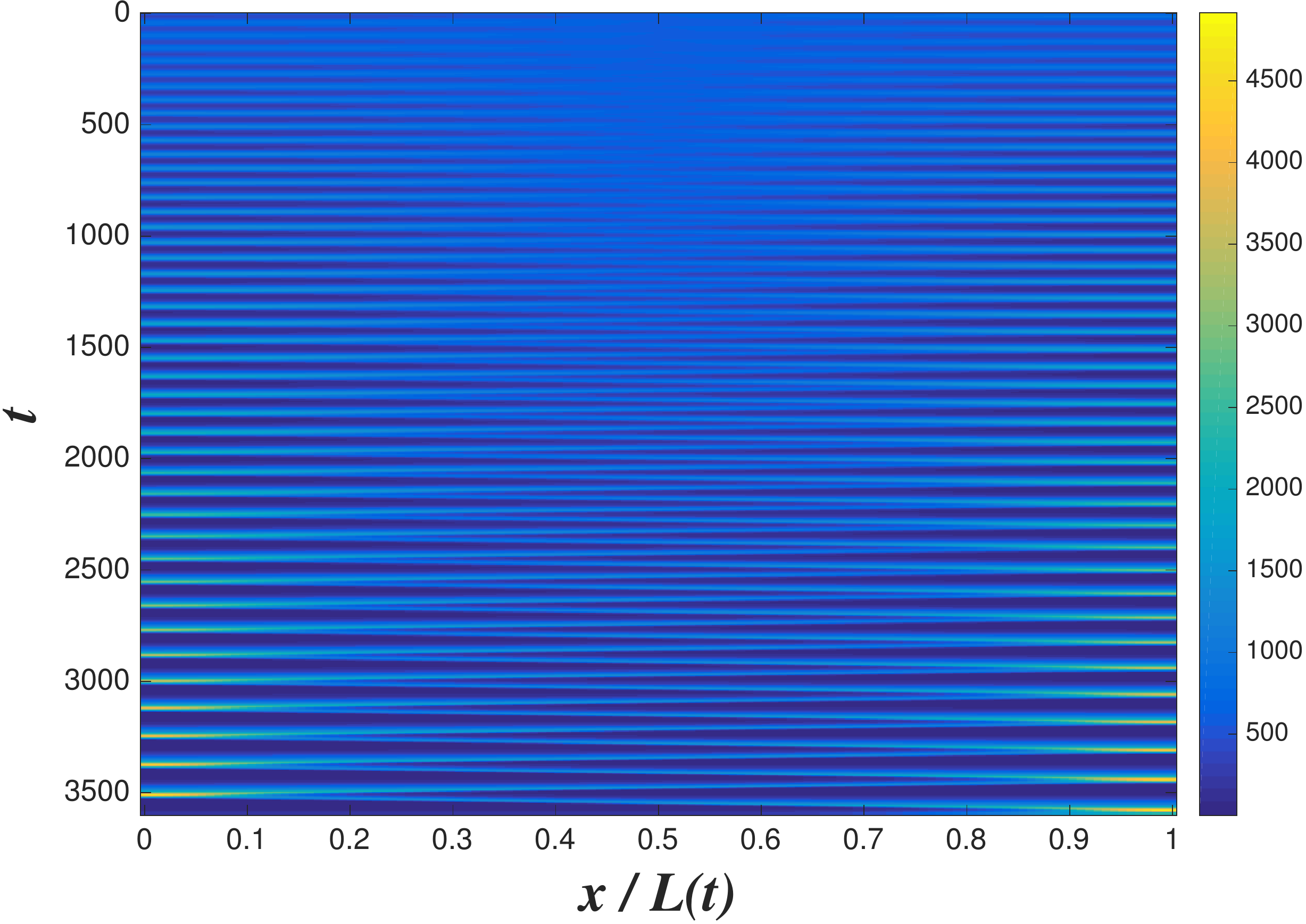}
 \caption{Top: A ``sloshing" pattern produced in an $L=2$ with reflecting boundary conditions. Bottom:  The adiabatic adjustment of the above ``sloshing" pattern in an exponentially growing system, with a doubling time of 30 mins. and an initial size $L=1.5$ The system quickly settles into the sloshing pattern, which for larger systems resembles the one-pulse traveling wave solution when the pulse is away from the end-walls.  The other parameters are the same as in the top panel of Fig. 2.}
 \end{figure}

\begin{acknowledgments}
This work was supported by the U.S. National Science Foundation Physics Frontier Center program grant no. PHY-1427654, the National Science Foundation Molecular and Cellular Biology (MCB) Division Grant MCB-1241332 and the U.S.-Israel Binational Science Foundation Grant no. 2015619. We gratefully acknowledge the hospitality of the Aspen Center for Physics, where this work was started.
\end{acknowledgments}

\end{document}